\documentclass[aps,preprint,superbib]{revtex4}%
\usepackage{amsfonts}
\usepackage{amsmath}
\usepackage{amssymb}
\usepackage{graphicx}%
\begin{document}
\title{A new comparison between solid-state thermionics and thermoelectrics}
\author{T. E. Humphrey}
\author{M. F. O'Dwyer}
\affiliation{School of Engineering Physics, University of Wollongong, Wollongong 2522, Australia}
\keywords{thermionics, thermoelectrics, efficiency, materials parameter}
\pacs{}

\begin{abstract}
It is shown that the equations for electrical current in solid-state
thermionic and thermoelectric devices converge for devices with a width equal
to the mean free path of electrons, yielding a common expression for intensive
electronic efficiency in the two types of devices. This result is used to
demonstrate that the materials parameters for thermionic and thermoelectric
refrigerators are equal, rather than differing by a multiplicative factor as
previously thought.

\end{abstract}
\maketitle

Solid-state thermionic devices may be distinguished from thermoelectric
devices according to whether electron transport is ballistic or diffusive
\cite{Nol01}. There is, however, little to distinguish the underlying
thermodynamics of the two types of devices, with both achieving reversibility
under the same conditions \cite{Hum02,Hum04}, and both being governed by the
same `materials parameter' \cite{Vin99,Sha99,Ulr01}. In previous work, a
proportionality factor of $F_{0}/F_{1/2}\sqrt{\pi}$ where $F_{n}$ is a Fermi
integral, was found between the two materials parameters \cite{Ulr01}. Here we
show that the equations for diffusive and ballistic transport converge for
devices with a width equal to the mean-free-path of electrons. This results in
a common equation for intensive electronic efficiency for both thermionic and
thermoelectric devices, and a proportionality factor between their material
parameters equal to unity. We explain the discrepancy between this and
previous work by pointing out an inconsistency in the energy dependency of the
relaxation time in two equations for conductivity used in \cite{Ulr01}.

Current in both thermionic and thermoelectric devices can be expressed as
\begin{equation}
J=\iiint j\left(  \mathbf{k}\right)  d\mathbf{k}%
\end{equation}
where $j\left(  \mathbf{k}\right)  \delta\mathbf{k}$ is the net
`energy-resolved' current of electrons flowing in the direction opposite to
the temperature gradient with momentum in the range $\delta\mathbf{k}$ around
$\mathbf{k}$. In thermionic devices with a width less than the mean-free-path,
most electrons travel ballistically from one reservoir to another. In this
case the energy resolved current density is given by (with dependence upon
\textbf{\thinspace}$\mathbf{k}$ implicit)
\begin{equation}
j^{b}\left(  \mathbf{k}\right)  \delta\mathbf{k}=qD_{r}\zeta v_{x}\Delta
f_{0}\delta\mathbf{k} \label{jb}%
\end{equation}
where $D_{r}\left(  \mathbf{k}\right)  $ is the density of states (DOS) in the
reservoirs, $\zeta\left(  \mathbf{k}\right)  $ is the probability that
electrons are transmitted between the reservoirs, $v_{x}\left(  \mathbf{k}%
\right)  $ is the velocity in the direction of transport, and $\Delta
f_{0}=\left[  f_{0}\left(  \mathbf{k,}\mu_{C},T_{C}\right)  -f_{0}\left(
\mathbf{k,}\mu_{H},T_{H}\right)  \right]  $ is the difference in the Fermi
occupation of states in the cold/hot reservoirs where%
\begin{equation}
f_{0}\left(  \mathbf{k,}\mu,T\right)  =\left[  1+\exp\left(  \frac{E\left(
\mathbf{k}\right)  -\mu}{kT}\right)  \right]  ^{-1}\text{.}%
\end{equation}
and where $\mu_{C/H}$ is the electrochemical potential and $T_{C/H}$ the
temperature of electrons at the hot/cold ends of the device.

Here we follow previous work \cite{Ulr01}, and assume the transmission
probability depends upon the total momentum of electrons rather than momentum
in the direction of transport only, allowing a direct comparison with
thermoelectrics in which the energy of mobile electrons is also restricted in
all three dimensions (this assumption is made implicitly in \cite{Ulr01},
between Eq. 1.1 and Eq. 1.3). The theoretical differences between thermionic
devices in which the transmission probability is a function of $\mathbf{k}$
and $k_{x}$ are explored in detail in other papers \cite{Hum04b,Vas04, ODw04}.

In thermoelectric devices the energy-resolved diffusive electron current
density may be obtained from the Boltzmann transport equation under the
relaxation time approximation, and can be written as \cite{Kit96}%
\begin{equation}
j^{d}\left(  \mathbf{k}\right)  \delta\mathbf{k}=qD_{l}\tau v_{x}^{2}%
\frac{df_{0}}{dx}\delta\mathbf{k} \label{jd}%
\end{equation}
where $D_{l}\left(  \mathbf{k}\right)  $ is the local DOS, $\tau\left(
\mathbf{k}\right)  =$ $\tau_{0}E\left(  \mathbf{k}\right)  ^{r}$ is the
relaxation time in the direction of transport.

In solid-state power generators and refrigerators with a width close to the
electronic mean-free-path it is expected that equations \ref{jb} and \ref{jd}
should yield the same results. To show this, we take the energy dependence of
the relaxation time to be $r=-1/2$, which corresponds to scattering that is
dominated by acoustic phonons, and results in a mean-free-path in the
direction of transport, $\lambda\equiv v_{x}\left(  \mathbf{k}\right)
\tau\left(  \mathbf{k}\right)  $, which is independent of energy \cite{Ulr01}.
We also note that $df_{0}\left(  x\right)  /dx\approx\left[  f_{0}\left(
x\right)  -f_{0}\left(  x+\delta x\right)  \right]  /\delta x$ when $\delta x$
is small, so that, for a piece of thermoelectric material $L=\lambda$ in
length, $df_{0}\left(  x\right)  /dx\approx\Delta f_{0}/\lambda$, and equation
\ref{jd} becomes%
\begin{equation}
j^{d}\left(  \mathbf{k}\right)  \delta\mathbf{k}=qD_{l}v_{x}\Delta f_{0}%
\delta\mathbf{k}.\label{jd1}%
\end{equation}
It can be seen that equations \ref{jb} and \ref{jd1} have the same form, where
the product of the DOS in the reservoirs and the transmission probability in
Eq. \ref{jb} plays the same role in determining the energy spectrum of
electrons which carry current as the local DOS does in Eq. \ref{jd1}. This
simple result provides an additional underpinning for Urich, Barnes and
Vining's observation in \cite{Ulr01} that thermionic and thermoelectric
devices refrigerate (or generate power) via the same underlying physical
mechanism. To show that there is no sharp transition in the behavior of a
solid-state power generator or refrigerator as its width changes from
$L<\lambda$ to $L>\lambda$, one can use the fact that the probability that an
electron can travel a distance $L$ without suffering a collision is
\cite{Ash76}
\begin{equation}
P=\exp\left(  -L/\lambda\right)
\end{equation}
to obtain an equation for energy-resolved current density useful in
solid-state power generators and refrigerators of length $L\approx\lambda$ as
\begin{equation}
j\left(  \mathbf{k}\right)  \delta\mathbf{k}=qv_{x}\Delta f_{0}\left\{
\frac{\lambda}{L}D_{l}\left[  1-P\right]  +D_{r}\zeta P\right\}
\delta\mathbf{k}\text{.}%
\end{equation}
which can easily be generalized to the case where $\lambda\rightarrow
\lambda\left(  \mathbf{k}\right)  $.

Equality between equations \ref{jb} and \ref{jd1} results in a common
expression for the electronic efficiency \cite{Hat73} (in thermionic devices)
and intensive efficiency \cite{Vin97, Sny03} (across a small section of
thermoelectric material) of
\begin{equation}
\Phi_{PG}^{b}=\left\vert VJ/\overset{\cdot}{Q}_{H}\right\vert \label{PGeffic}%
\end{equation}
for a solid-state power generator and
\begin{equation}
\Phi_{R}^{b}=\left\vert \overset{\cdot}{Q}_{C}/VJ\right\vert \label{Reffic}%
\end{equation}
for a solid-state refrigerator, where the heat flux density in the cold/hot
reservoir of a thermionic device, or at the cold/hot ends of the small section
of thermoelectric is given by
\begin{equation}
\overset{\cdot}{Q}_{C/H}=\mp\iiint\left[  E\left(  \mathbf{k}\right)
-\mu_{C/H}\right]  \frac{j\left(  \mathbf{k}\right)  }{q}d\mathbf{k}%
\label{heat}%
\end{equation}
where the -/+ refers to the cold/hot case, and it is assumed that the
temperature gradient, electric field ($\varepsilon=V/L$), and current have no
components in the $y$ and $z$ dimensions. It is important to point out that
the Eq. \ref{heat} considers only heat carried by electrons. If phonon
mediated heat leaks are also considered, then two extra terms should be added
to Eq. \ref{heat}, the first being $\pm\Delta T\kappa_{l}/L$, where
$\kappa_{l}$ is the thermal conductivity of the lattice, while the second
accounts for `Joule' heat released by electrons which diffuse between the hot
and cold reservoirs, part of which is then carried by phonons to each end of
the device. Shakouri et al. \cite{Sha98} have shown that in the limit that
$L\gg\lambda$ then heat exactly half of this heat, $VJ$, is deposited in the
hot and cold reservoirs, while if $L\ll\lambda$ this heat is entirely
deposited in the reservoir recieving the net current of electrons.

Finally, it can be shown that the above results yield the same expression for
the materials parameter in thermionic and thermoelectric refrigerators.
Assuming a dispersion relation of $E=\hbar^{2}\mathbf{k}^{2}/2m^{\ast}$, and
three-dimensional reservoirs so that $D=1/\left(  2\pi\right)  ^{3}$, Eq.
\ref{jb} may be used to obtain the same thermionic materials parameter as
found in \cite{Ulr01}, expressed in terms of the mean-free-path $\lambda$ as%
\begin{equation}
\beta_{TI}=\frac{\lambda}{\kappa_{l}}\frac{4\pi m^{\ast}k}{h^{3}}\left(
kT_{C}\right)  ^{2}%
\end{equation}
where $\kappa_{l}$ is the thermal conductivity of the lattice. The
thermoelectric materials parameter may be obtained from \cite{Ulr01}%
\begin{equation}
\beta=\frac{\sigma T}{\kappa_{l}}\frac{k^{2}}{q^{2}}F_{0}\left(  \eta\right)
^{-1}%
\end{equation}
where $\sigma$ is the electrical conductivity and $F_{n}$ is a Fermi integral,
given by
\begin{equation}
F_{n}\left(  \eta\right)  =\frac{1}{\Gamma\left(  n+1\right)  }\int
_{0}^{\infty}\frac{\epsilon^{n}}{\exp\left(  \epsilon-\eta\right)
+1}d\epsilon
\end{equation}
where the reduced Fermi energy is $\eta=\left(  \mu-E^{\prime}\right)
/kT_{C}$ and $E^{\prime}$ is the height of the barrier in thermionic devices
(so that $\zeta=0$ for $E<E^{\prime}$, $\zeta=1$ for $E>E^{\prime}$) or the
conduction band edge in a piece of thermoelectric. Using the same dispersion
relation and density of states as for thermionic devices we can write an
equation for conductivity in thermoelectric materials as
\begin{align}
\sigma &  \equiv\frac{qJ^{d}}{d\mu/dx}=\frac{4\pi m^{\ast}q^{2}\lambda}{h^{3}%
}\int E\frac{df_{0}\left(  E\right)  }{dE}dE\\
&  =\lambda\frac{4\pi m^{\ast}k}{h^{3}}q^{2}TF_{0}\left(  \eta\right)
\end{align}
where we have used the fact that $df_{0}/dx=\left(  d\mu/dx\right)  \left(
df_{0}/d\mu\right)  $ for $dT/dx=0$, that $df_{0}/d\mu=-df_{0}/dE$, and have
integrated by parts to obtain the final line. It can be seen by substitution
that this yields $\beta=\beta_{TI}$ if $T=T_{C}$.

In \cite{Ulr01} the thermionic and thermoelectric materials parameters were
expressed in terms of the mobility of electrons. In order to do this for the
thermoelectric materials parameter, two different expressions for conductivity
were equated:%
\begin{equation}
\sigma=-q^{2}\int v_{x}^{2}\tau D_{l}\frac{df_{0}}{dE}dE
\end{equation}
where $\tau=\tau_{0}E^{-1/2}$, and
\begin{align}
\sigma &  =nq\mu=q\mu\int D_{l}f_{0}dE\\
&  =-\frac{q^{2}\tau}{m^{\ast}}\int D_{l}E\frac{df_{0}}{dE}\text{.}%
\end{align}
Given that $v_{x}^{2}\propto E$ and $\mu\equiv q\tau/m^{\ast}$, it can be seen
that the two equations do not give the same result for conductivity as the
relaxation time is assumed to be proportional to $E^{-1/2}$ in the first and
independent of energy in the second, and this difficulty results in the factor
of $F_{0}/F_{1/2}\sqrt{\pi}$ identified in \cite{Ulr01}.

In summary, it has been shown that solid-state thermionic and thermoelectric
devices with a width close to the mean-free-path of electrons share a common
equation for electrical current density and electronic efficiency, and that
their `material parameters' for refrigeration are equal.

\bibliographystyle{apsrev}
\bibliography{Tammysbib}

\begin{thebibliography}{15}
\expandafter\ifx\csname natexlab\endcsname\relax\def\natexlab#1{#1}\fi
\expandafter\ifx\csname bibnamefont\endcsname\relax
  \def\bibnamefont#1{#1}\fi
\expandafter\ifx\csname bibfnamefont\endcsname\relax
  \def\bibfnamefont#1{#1}\fi
\expandafter\ifx\csname citenamefont\endcsname\relax
  \def\citenamefont#1{#1}\fi
\expandafter\ifx\csname url\endcsname\relax
  \def\url#1{\texttt{#1}}\fi
\expandafter\ifx\csname urlprefix\endcsname\relax\def\urlprefix{URL }\fi
\providecommand{\bibinfo}[2]{#2}
\providecommand{\eprint}[2][]{\url{#2}}

\bibitem[{\citenamefont{Nolas et~al.}(2001)\citenamefont{Nolas, Sharp, and
  Goldsmid}}]{Nol01}
\bibinfo{author}{\bibfnamefont{G.~S.} \bibnamefont{Nolas}},
  \bibinfo{author}{\bibfnamefont{J.}~\bibnamefont{Sharp}}, \bibnamefont{and}
  \bibinfo{author}{\bibfnamefont{H.~J.} \bibnamefont{Goldsmid}},
  \emph{\bibinfo{title}{Thermoelectrics Basic Principles and New Materials
  Developments}} (\bibinfo{publisher}{Springer}, \bibinfo{address}{Berlin},
  \bibinfo{year}{2001}).

\bibitem[{\citenamefont{Humphrey and Linke}(2005)}]{Hum04}
\bibinfo{author}{\bibfnamefont{T.~E.} \bibnamefont{Humphrey}} \bibnamefont{and}
  \bibinfo{author}{\bibfnamefont{H.}~\bibnamefont{Linke}},
  \bibinfo{journal}{Phys. Rev. Lett. (In Press) condmat/0407509}
  (\bibinfo{year}{2005}).

\bibitem[{\citenamefont{Humphrey et~al.}(2002)\citenamefont{Humphrey, Newbury,
  Taylor, and Linke}}]{Hum02}
\bibinfo{author}{\bibfnamefont{T.~E.} \bibnamefont{Humphrey}},
  \bibinfo{author}{\bibfnamefont{R.}~\bibnamefont{Newbury}},
  \bibinfo{author}{\bibfnamefont{R.~P.} \bibnamefont{Taylor}},
  \bibnamefont{and} \bibinfo{author}{\bibfnamefont{H.}~\bibnamefont{Linke}},
  \bibinfo{journal}{Phys. Rev. Lett.} \textbf{\bibinfo{volume}{89}},
  \bibinfo{pages}{116801} (\bibinfo{year}{2002}).

\bibitem[{\citenamefont{Shakouri and LaBounty}(1999)}]{Sha99}
\bibinfo{author}{\bibfnamefont{A.}~\bibnamefont{Shakouri}} \bibnamefont{and}
  \bibinfo{author}{\bibfnamefont{C.}~\bibnamefont{LaBounty}},
  \bibinfo{journal}{Proceedings of the 18th International Conference on
  Thermoelectrics, Baltimore} p.~\bibinfo{pages}{35} (\bibinfo{year}{1999}).

\bibitem[{\citenamefont{Ulrich et~al.}(2001)\citenamefont{Ulrich, Barnes, and
  Vining}}]{Ulr01}
\bibinfo{author}{\bibfnamefont{M.~D.} \bibnamefont{Ulrich}},
  \bibinfo{author}{\bibfnamefont{P.~A.} \bibnamefont{Barnes}},
  \bibnamefont{and} \bibinfo{author}{\bibfnamefont{C.~B.}
  \bibnamefont{Vining}}, \bibinfo{journal}{J. Appl. Phys.}
  \textbf{\bibinfo{volume}{90}}, \bibinfo{pages}{1625} (\bibinfo{year}{2001}).

\bibitem[{\citenamefont{Vining and Mahan}(1999)}]{Vin99}
\bibinfo{author}{\bibfnamefont{C.~B.} \bibnamefont{Vining}} \bibnamefont{and}
  \bibinfo{author}{\bibfnamefont{G.~D.} \bibnamefont{Mahan}},
  \bibinfo{journal}{Appl. Phys. Lett.} \textbf{\bibinfo{volume}{86}},
  \bibinfo{pages}{6852} (\bibinfo{year}{1999}).

\bibitem[{\citenamefont{Humphrey et~al.}(2004)\citenamefont{Humphrey, O'Dwyer,
  and Linke}}]{Hum04b}
\bibinfo{author}{\bibfnamefont{T.~E.} \bibnamefont{Humphrey}},
  \bibinfo{author}{\bibfnamefont{M.}~\bibnamefont{O'Dwyer}}, \bibnamefont{and}
  \bibinfo{author}{\bibfnamefont{H.}~\bibnamefont{Linke}},
  \bibinfo{journal}{Submitted to J. Phys. D condmat/0401377}
  (\bibinfo{year}{2004}).

\bibitem[{\citenamefont{O'Dwyer et~al.}(2005)\citenamefont{O'Dwyer, Humphrey,
  Lewis, and Zhang}}]{ODw04}
\bibinfo{author}{\bibfnamefont{M.}~\bibnamefont{O'Dwyer}},
  \bibinfo{author}{\bibfnamefont{T.~E.} \bibnamefont{Humphrey}},
  \bibinfo{author}{\bibfnamefont{R.}~\bibnamefont{Lewis}}, \bibnamefont{and}
  \bibinfo{author}{\bibfnamefont{C.}~\bibnamefont{Zhang}}, \bibinfo{journal}{In
  preparation}  (\bibinfo{year}{2005}).

\bibitem[{\citenamefont{Vashaee and Shakouri}(2004)}]{Vas04}
\bibinfo{author}{\bibfnamefont{D.}~\bibnamefont{Vashaee}} \bibnamefont{and}
  \bibinfo{author}{\bibfnamefont{A.}~\bibnamefont{Shakouri}},
  \bibinfo{journal}{Phys. Rev. Lett.} \textbf{\bibinfo{volume}{92}},
  \bibinfo{pages}{106103} (\bibinfo{year}{2004}).

\bibitem[{\citenamefont{Kittel}(1996)}]{Kit96}
\bibinfo{author}{\bibfnamefont{C.}~\bibnamefont{Kittel}},
  \emph{\bibinfo{title}{Introduction to Solid State Physics}}
  (\bibinfo{publisher}{John Wiley and Sons}, \bibinfo{address}{New York},
  \bibinfo{year}{1996}).

\bibitem[{\citenamefont{Ashcroft and Mermin}(1976)}]{Ash76}
\bibinfo{author}{\bibfnamefont{N.~W.} \bibnamefont{Ashcroft}} \bibnamefont{and}
  \bibinfo{author}{\bibfnamefont{N.~D.} \bibnamefont{Mermin}},
  \emph{\bibinfo{title}{Solid State Physics}} (\bibinfo{publisher}{Saunders
  College Publishing}, \bibinfo{address}{Orlando, Florida},
  \bibinfo{year}{1976}), p. \bibinfo{pages}{249}.

\bibitem[{\citenamefont{Hatsopoulos and Gyftopoulos}(1973)}]{Hat73}
\bibinfo{author}{\bibfnamefont{G.~N.} \bibnamefont{Hatsopoulos}}
  \bibnamefont{and} \bibinfo{author}{\bibfnamefont{E.~P.}
  \bibnamefont{Gyftopoulos}}, \emph{\bibinfo{title}{Thermionic Energy
  Conversion - Vol. 1: Processes and devices}} (\bibinfo{publisher}{MIT Press},
  \bibinfo{address}{Massachusetts, U.S.A.}, \bibinfo{year}{1973}).

\bibitem[{\citenamefont{Snyder and Ursell}(2003)}]{Sny03}
\bibinfo{author}{\bibfnamefont{G.~J.} \bibnamefont{Snyder}} \bibnamefont{and}
  \bibinfo{author}{\bibfnamefont{T.~S.} \bibnamefont{Ursell}},
  \bibinfo{journal}{Phys. Rev. Lett.} \textbf{\bibinfo{volume}{91}},
  \bibinfo{pages}{148301} (\bibinfo{year}{2003}).

\bibitem[{\citenamefont{Vining}(1997)}]{Vin97}
\bibinfo{author}{\bibfnamefont{C.}~\bibnamefont{Vining}},
  \emph{\bibinfo{title}{Symposium on Thermoelectric Materials - New Directions
  and Approaches}} (\bibinfo{publisher}{Materials Research Society},
  \bibinfo{address}{Pittsburgh}, \bibinfo{year}{1997}), p.~\bibinfo{pages}{3}.

\bibitem[{\citenamefont{Shakouri et~al.}(1998)\citenamefont{Shakouri, Lee,
  Smith, Narayanamurti, and Bowers}}]{Sha98}
\bibinfo{author}{\bibfnamefont{A.}~\bibnamefont{Shakouri}},
  \bibinfo{author}{\bibfnamefont{E.~Y.} \bibnamefont{Lee}},
  \bibinfo{author}{\bibfnamefont{D.~L.} \bibnamefont{Smith}},
  \bibinfo{author}{\bibfnamefont{V.}~\bibnamefont{Narayanamurti}},
  \bibnamefont{and} \bibinfo{author}{\bibfnamefont{J.~E.}
  \bibnamefont{Bowers}}, \bibinfo{journal}{Microscale Thermophysical
  Engineering} \textbf{\bibinfo{volume}{2}}, \bibinfo{pages}{37}
  (\bibinfo{year}{1998}).

\end{thebibliography}

\end{document}